\documentclass[amsmath,prb,twocolumn,showpacs,floatfix]{revtex4}
\usepackage{graphicx}
\usepackage{amssymb}
\newfont{\Sc}{eusm10}

\def\ScH{\mbox{\Sc \symbol{72}}}

\begin{document}
\title{Study of the Magnetic Film Materials by Horizontal Scanning Mode for the Magnetic Force Microscopy in Magnetostatic and ac Regimes.}

\author{Artorix de la Cruz de O\~{n}a${}^{1,2}$ and Nibaldo Alvarez-Moraga${}^{2}$}
\affiliation{${}^1$Department of Mathematics and Statistics,
Concordia University, Montr\'eal, Qu\'ebec, H3G 1M8, Canada
\\
{ \it  ${}^2$ \small \ Autonomous Center of theoretical Physics and
Applied Mathematics,} {\small \it 11996 Jubinville $\#$ 7,
Montr\'eal-Nord (Qu\'ebec) H1G 3T2, Canada} }

\date{May 3, 2006}
%
\vspace{-15mm}
\begin{abstract}
The magnetic force microscopy inverse problem for the case of
horizontal scanning of a tip on a linear magnetic film is
introduced. We show the possibility to recover the magnetic
permeability of the material from the experimental data by using the
Hankel (Fourier-Bessel) transform inverse method (HIM). This method
is applied to the case of a layered slab film as well. The inverse
problem related to the ac MFM is introduced.
\end{abstract}
\maketitle
\section{Introduction}

In the theory of inverse magnetic force microscopy (MFM), the
recovering of the magnetic permeability is based on the interaction
of a tip with the stray field of a magnetic material. So far the
inverse problem has been solved in magnetostatic considerations for
different cases such as spherical\cite{coffeyprb1},
seminfinite\cite{coffeyinverse} and slab\cite{delacruzPhysB04}
geometries. In order to measure the magnetostatic interaction force,
the direction of the MFM tip movement has been always assumed to be
perpendicular to the surface of the sample.

This vertical movement mode of the MFM entails a mathematical
procedure based on the Laplace transform inverse method
(LIM)\cite{badiaprb99,badiaprb01,coffeyotherinver,coffeyprl,coffeyprb98}
applied to experimental data. Solutions of the LIM are not unique
since the numerical Laplace transform inversion is intrinsically
unstable\cite{badiaprb01,coffeyprb0}. In other words, small
variations in the initial conditions or the noise present in the
experimental data cause large variations of results.

In this paper, we assume lateral displacements of the tip which
allow us to consider the use of the Hankel transform inverse method
(HIM)\cite{artobadia}, an alternative approach to LIM, for
recovering the magnetic permeability $\mu_{r}$. We first present the
inverse problem related to the MFM for a finite magnetic slab. We
extend the proposed Hankel inverse algorithm to the case of a finite
layered film and recover the magnetic properties as well as the
geometrical characteristics of each layer.

Finally, we introduce the HIM for the $ac$ MFM regime to study a
magnetic slab.
%
\section{Slab film magnetostatic inverse problem}
Let us consider the tip as a magnetic point dipole $\vec{m}$ with
components $(0,0,m_{z})$ placed at the position $(0,0,a)$  outside a
magnetic slab film of thickness $b$. The film is located at $z=0$
parallel to the $xy$ plane. We assume the relative magnetic
permeability $\mu_{r}=\mu/\mu_{0}$ of the material to be constant,
where $\mu_{0}$ is the magnetic permeability in the vacuum. The
scalar potential $\phi$ satisfies the Poisson equation
\begin{equation}
\label{eq:poisson}
\nabla^{2}\phi=\vec{m}\cdot\vec{\nabla}\delta(x)\,\delta(y)\,\delta(z-a).
\end{equation}
The general solution $\phi$ of the Poisson equation is given by a
sum of a particular solution $\phi_{1}$ and a homogeneous solution
$\phi_{2}$. This decomposition of $\phi$ corresponds to the
decomposition of the induction $\vec{B}(\vec{r})$ (where
$\vec{B}(\vec{r})=-\mu_{0}\vec{\nabla}\phi$), into the sum
$\vec{B}(\vec{r})=\vec{B}_{1}(\vec{r})+\vec{B}_{2}(\vec{r})$, where
$\vec{B}_{1}(\vec{r})$ represents the direct magnetic field
induction due to the dipole, and $\vec{B}_{2}(\vec{r})$ is the
reflected magnetic induction.

For the dipole $\vec{m}$, we have
\begin{equation}
\label{eq:potential}
\phi_{1}(\vec{r})=\,\frac{\mu_{0}}{4\pi}\,\frac{\vec{m}\cdot\vec{r}}{{|\vec{r}_{1}|^{3}}},
\end{equation}
where $\vec{r}_1 \equiv(x,y,z-a)$.

Now, we expand $\phi_{1}(\vec{r})$ in terms of the Bessel functions.
For that we use the following identity
\begin{eqnarray}
\label{eq:NAbla}
\frac{1}{\sqrt{\rho^{2}+z^{2}}}=\int_{0}^{\infty}dk\,e^{-k|z|}\,J_{0}(k\rho).
\end{eqnarray}
Here and throughout, $J_{\ell}$ denotes the Bessel function of
$\ell^{th}$ order of the first kind and $\rho^{2}=x^{2}+y^{2}$ .
Eq.(\ref{eq:NAbla}) and the recurrence relations of Bessel functions
permit us to express the $z$-component of $\phi_{1}(\vec{r})$ from
Eq.(\ref{eq:potential}) in the following\cite{delacruzPhysB04}
compact form,

\begin{equation}
\label{eq:Nabla}
\phi_{z1}=\frac{m_{z}}{4\pi}\int_{0}^{\infty}dk\,e^{-k(z-a)}\,k\,J_{0}(k\rho).
\end{equation}
The distribution of the magnetic field induction $\vec{B}(\vec{r})$
is given by
\begin{eqnarray}
\label{eq:NABla}
\vec{B}(\vec{r})=\left\{
\begin{array}{l}
\vec{B}_{1}(\vec{r})+\vec{B}_{2}(\vec{r})\qquad z > 0 \\
\vec{B}_{3}(\vec{r})\qquad\qquad -b < z < 0 \\
\vec{B}_{4}(\vec{r})\qquad\qquad\qquad  z < -b \; ,\\
\end{array}
\right.
\end{eqnarray}
where $\vec{B}_{3}(\vec{r})$ and $\vec{B}_{4}(\vec{r})$ are the
penetrating and transmitted magnetic fields respectively.

The following governing elliptic differential equations hold for
$\vec{B}(\vec{r})$

\begin{equation}
\label{eq:Nablaa} \nabla^{2}\vec{B}_{i}(\vec{r})=0,
\,\,(i=1,\ldots,4).
\end{equation}

Then, analogous to Eq.(\ref{eq:Nabla}), we can write the
$z$-component of the scalar potentials\cite{delacruzPhysB04} as
follows:
\begin{eqnarray}
\label{eq:NaBLA} \phi_{z2} &=& \frac{m_{z}}{4\pi}\int_{0}^{\infty}dk
k I_{z2}(k) e^{-k(a+z)} J_{0}(k \rho), \nonumber \\
\phi_{z3} &=& \frac{m_{z}}{4\pi}\int_{0}^{\infty}dk k [I_{z3}^{+}
e^{k z}+I_{z3}^{-} e^{-k z}] e^{-ka} J_{0}(k \rho),\nonumber\\
\phi_{z4} &=& \frac{m_{z}}{4\pi}\int_{0}^{\infty}dk k I^{+}_{z4}(k)
e^{-k(a-z-b)} J_{0}(k \rho).
\end{eqnarray}
The coefficients $I_{z_{i}}(k)$ can be obtained if we impose the
continuity boundary conditions on $\phi_{z_{i}}$ and $\mu\,\partial
\phi_{z_{i}}/\partial z$ at the interfaces ($z=0, z=-b$). Notice
that we do not consider the solutions which are divergent as
$z\rightarrow\pm\infty$ for the fields $B_{z3}$ and $B_{z4}$.
We are interested in the coefficient $I_{z_{2}}(k)$ since $B_{z2}$
is the only magnetic field that interacts with the tip. For this
coefficient we have from Eq.(\ref{eq:NaBLA})
\begin{eqnarray}
\label{eq:Coefficien} I_{z2}(\mu_{r},k) &=&
\Delta_{\mu}\,(e^{2bk}-1)\,(1-\mu_{r}^{2}),
\end{eqnarray}
where $\Delta_{\mu}\equiv
(e^{2bk}(\mu_{r}+1)^{2}-(\mu_{r}-1)^2)^{-1}$.
The expressions for the incident, reflected, penetrating and
transmitted fields respectively are
\begin{eqnarray}
\label{eq:NonLOCAL} B_{z_{1}} &=&
\frac{\mu_{0}\,m_{z}}{4\pi}\int_{0}^{\infty}dk\,k^2\,e^{-k(a-z)}J_{0}(k\rho),
\end{eqnarray}
\begin{eqnarray}
\label{eq:Lola} B_{z_{2}} &=&
\frac{\mu_{0}\,m_{z}}{4\pi}\int_{0}^{\infty}dk\,k^2\,I_{z_{2}}(\mu_{r},k)\,e^{-k(a+z)}J_{0}(k\rho),
\end{eqnarray}
\begin{eqnarray}
\label{eq:Lola1}
B_{z_{3}} &=& \frac{\mu\,m_{z}}{4\pi}\int_{0}^{\infty}dk k^2[I_{z_{3}}^{+}(\mu_{r},k) e^{k z}+I_{z_{3}}^{-}(\mu_{r},k) e^{-k z}]\nonumber \\
&\times& \, e^{-ka} J_{0}(k\rho),
\end{eqnarray}
\begin{eqnarray}
\label{eq:Lola2} B_{z_{4}} &=&
\frac{\mu_{0}\,m_{z}}{4\pi}\int_{0}^{\infty}dk\,k^2\,I_{z_{4}}(\mu_{r},k)\,e^{-k(a-z-b)}J_{0}(k\rho).
\end{eqnarray}

The interaction force between the tip and the slab is given by
\begin{equation}
\label{eq:NForce}
F_{z}(a)=-\frac{\partial}{\partial a}\,(-\frac{1}{2}\:\vec{m}\cdot
\vec{B}_{2}),
\end{equation}
and it can be rewritten using Eq.(\ref{eq:Lola}) as follows
\begin{equation}
\label{eq:Force}
F_{z}(a)=\frac{\mu_{0}
m^{2}_{z}}{4\pi}\int^{\infty}_{0}\,dk\,k^{3}\,I_{z2}(\mu_{r},k)\,e^{-2a\,k}\,J_{0}(k\rho).
\end{equation}

Let us assume that we need to recover the unknown $\mu_{r}$ from the
MFM data (i.e. the force $F_{z}$). Consider a horizontal
displacement of the tip such that the distance $a$ between the tip
and the magnetic film remains constant, $a=a_{0}$.

Recalling the definition of the Hankel transform operator
\begin{eqnarray}
\label{eq:Hankel}
\ScH_{0}[f(k)](\rho)=\int kJ_{0}(k\rho)f(k)dk,
\end{eqnarray}
we can write Eq.(\ref{eq:Force}) as follows
\begin{equation}
\label{eq:cOFficient}
F_{z}(a_0,\rho)=\frac{\mu_{0}\,m^{2}_{z}}{4\pi}\ScH_{0}[k^{2}
I_{z2}(\mu_{r},k)\,e^{-2k a_0}].
\end{equation}
The data inversion process involved in our algorithm reduces to a
simpler operation, the so-called Hankel transform inversion. The
forward and inverse transformations have the same operational
form\cite{artobadia} $\ScH_{0}^{-1}=\ScH_{0}$.

Let us now describe the algorithm for finding $\mu_{r}$ starting
with the experimental force data. The mathematical inversion should
be performed for the pairs $[\rho_{i},F_{z}(a_{0},\rho_{i})]$ where
$\rho$ is the Hankel transformed variable. We apply the inverse
operator $\ScH_{0}^{-1}$ to both sides of Eq.(\ref{eq:cOFficient}).

Consider the quantity $\ScH_{0}^{-1}[F_{z}(a_{0},\rho)](k)$ to be
known for some collection of values of the wave-number $k$. Using
these data it is possible to determine the pair $(\mu_{r},b)$ or
$(\mu_{r},I_{z2})$ as follows. Assuming $k$ as a parameter, we can
take any nontrivial pair $k_{1}\neq k_{2}$ and write the following
nonlinear system of equations:

\begin{figure}[!tbp]
\centerline{
\includegraphics[width=3.2in]{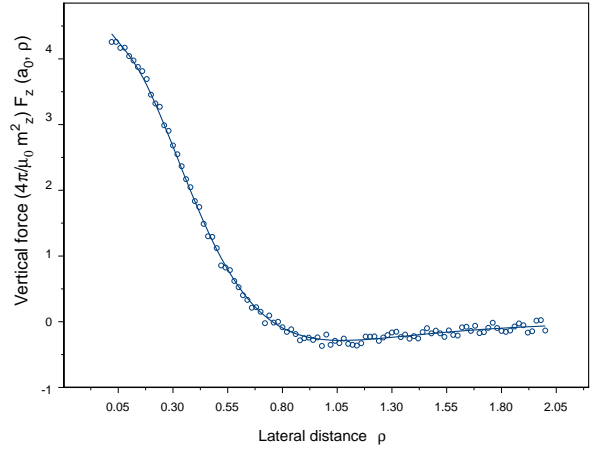}
} \caption{Plot of the simulated force $F_{z}(a_{0},\rho)$ used in
this work. A noise corrupted data have been added.} \label{figure1}
\end{figure}

\begin{eqnarray}
\label{eq:SystemHankel}
\nonumber
\frac{\mathcal{T}(a_{0},k_{1})}{k^{2}_{1}}\;\ScH_{0}^{-1}[F_{z}(a_0,\rho)](k_{1})-{
I}_{z2}(\mu_{r},k_{1})
&=& 0\\
\frac{\mathcal{T}(a_{0},k_{2})}{k^{2}_{2}}\;\ScH_{0}^{-1}[F_{z}(a_0,\rho)](k_{2})-{I}_{z2}(\mu_{r},k_{2})
&=& 0,
\end{eqnarray}
where $\mathcal{T}(a_{0},k)\equiv 4\pi {\rm e}^{2k
a_0}/\mu_{0}m^{2}_{z}$.

\begin{figure}[!tbp]
\centerline{
\includegraphics[width=3.2in]{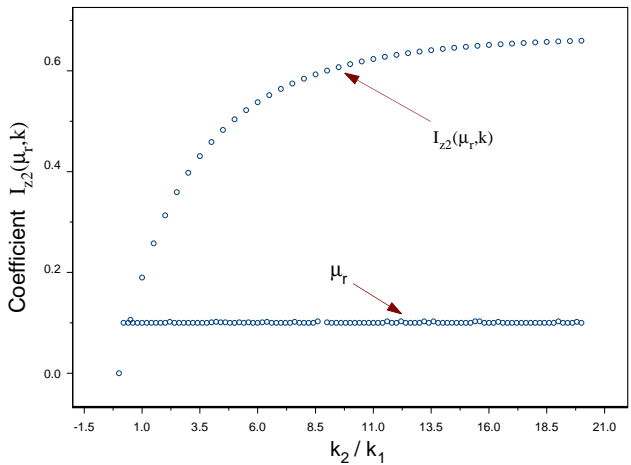}
} \caption{The magnetic permeability $\mu_{r}$ and the coefficient
$I_{z2}$ recovered by using the Hankel transform technique.}
\label{figure2}
\end{figure}

The system (\ref{eq:SystemHankel}) allows us to obtain the quantity
$\mu_{r}$ from the force experimental data. To prove that the
described method gives the value of $\mu_{r}$, we perform a
simulation\cite{badiaprb99,artobadia} with a set of faked MFM data
with $a_{0}=0.435,\, b=0.2,\, \rho=0.2$ and $\mu_{r}=0.1$. Here and
throughout the paper the units are measured in $\mu m$. The
corresponding graph is shown in the Fig.1.

Fig.2 displays the results of the recovering of $\mu_{r}$ and the
coefficient $I_{z2}(\mu_{r},k)$. The results agree with the
theoretical value of the parameter $\mu_{r}$ used in the forward
problem shown in Fig.1, in a large range of values of the
wave-number $k$.
\section{Layered film magnetostatic inverse problem }
The inverse problem for the slab can be generalized in order to
study the magnetic multilayered films.

Let us consider a finite slab film that geometrically can be viewed
as a collection of $M$ layers. Denote the $z$-coordinate of the
$j$-th layer by $z_{j}$. We say that the lowest layer is designated
as layer 1, while layer $M$ is the top of the slab and $z_{M}=0$.
The slab has a total thickness $z_{1}=-b$. We assume a constant
magnetic permeabilities $\mu_{rj}$ $(j=1,...,M)$ to be different for
each layer.

The distribution of the magnetic field $\vec{B}(\vec{r})$ can be
written as
\begin{eqnarray}
\label{eq:cOFFICIent}
&\vec{B}_{1}(\vec{r})+\vec{B}_{2}(\vec{r})& z>0, \nonumber\\
&\vec{B}_{3}(\vec{r})& z_{j}<z<z_{j+1}, \\
&\vec{B}_{4}(\vec{r})& z<z_{1}.\nonumber
\end{eqnarray}

The elliptic differential equations
$\nabla^{2}\vec{B}_{i}=0\,(i=1,\ldots,4)$ hold. The $z$-components
of $\vec{B}_{2}, \vec{B}_{3}$ and $\vec{B}_{4}$ are given by
\begin{eqnarray}
\label{eq:cofFICIEnt}
B_{z_{2}} &=& \frac{\mu_{0} m_{z}}{4\pi}\int_{0}^{\infty}dk k^2 I^{(M+1)}_{z_{2}}(k) e^{-k(a+z)}J_{0}(k\rho), \\
B_{z_{3}} &=& \frac{\mu m_{z}}{4\pi}\int_{0}^{\infty}dk k^2 [I_{z_{3}}^{j+}(k)e^{k z}+I_{z_{3}}^{j-}(k) e^{-k z}]\nonumber \\
&\times& e^{-ka} J_{0}(k\rho), \\
B_{z_{4}} &=& \frac{\mu_{0} m_{z}}{4\pi}\int_{0}^{\infty}dk k^{2}
I_{z_{4}}(k) e^{-k(a-z-z_{1})}J_{0}(k\rho).
\end{eqnarray}
The coefficients $I_{z_{i}}(k)$ may be obtained by imposing
continuity boundary conditions at the planar interfaces $z_{M}=0$
and $z=z_{j}$ on $B_{z_{i}}$ in a way similar to that described in
section II.

Taking into account Eqs.(\ref{eq:NForce}) and (\ref{eq:cofFICIEnt}),
we can write the expression for the force between the tip and the
layered film as follows
\begin{equation}
\label{eq:MAGNetic} F_{z}(a)=\frac{\mu_{0}
m^{2}_{z}}{4\pi}\int^{\infty}_{0}dkk^{3}I^{(M+1)}_{z2}e^{-2ak}J_{0}(k\rho).
\end{equation}

Notice that $I^{(M+1)}_{z2}$ contains all the geometrical and
physical information related to the layered film. This coefficient
is a function of the permeability and position of each layer, i.e.
$I^{(M+1)}_{z2}=I^{(M+1)}_{z2}(k,z_{j},\mu_{rj})$.

Taking into account the definition (\ref{eq:Hankel}) of the Hankel
transform $\ScH_{0}$, we apply its inverse $\ScH_{0}^{-1}$ to both
sides of Eq.(\ref{eq:MAGNetic}). For $M$ couples of values
$k_{j+1}\neq k_{j}$ we get the following nonlinear system, which is
analogous to (\ref{eq:SystemHankel}):
\begin{eqnarray}
\label{eq:foLLOWING}
\frac{\mathcal{T}(a_{0},k_{M+1})}{k_{M+1}^2}\ScH_{0}^{-1}[F_{z}(a_0,\rho)](k_{M+1})-I^{(M+1)}_{z2}(k_{M+1})
&=& 0 \nonumber \\
\cdots\\
\frac{\mathcal{T}(a_{0},k_{j})}{k_{j}^2}\ScH_{0}^{-1}[F_{z}(a_0,\rho)](k_{j})-I^{(M+1)}_{z2}(k_{j})
&=& 0. \nonumber
\end{eqnarray}
\begin{figure}[!tbp]
\centerline{
\includegraphics[width=3.2in]{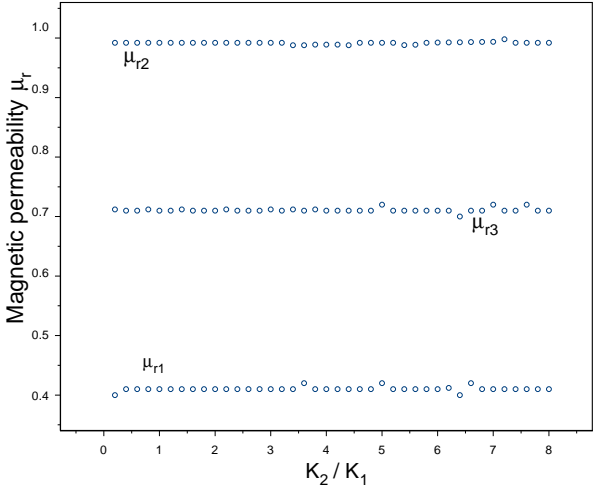}
} \caption{Recovery of the magnetic permeabilities
$\mu_{r1},\mu_{r2}$ and $\mu_{r3}$ for a slab with $M$=3. The
theoretical values assumed for $\mu_{rj}$ are $0.41, 0.9, 0.71$,
respectively.} \label{figure4}
\end{figure}

In order to get the force noise corrupted data, we have followed the
simulation procedure used in section II for the slab geometry. Fig.3
shows the result of applying the described inverse method to a film
having three layers.

Notice that the magnetic layered film-tip interaction problem
involves an attenuating magnetic field penetration\cite{coffeyprb0}.
If the experimental data of the force reflect this fact, then better
results in the inverse procedure for layers situated next to the top
of the slab $(z=0)$ can be archived.
\section{$ac$ Regime for a slab film}
It is well-known that one achieves better results using the MFM if
one considers the $ac$ regime rather than a magnetostatic
interaction between the tip and the film\cite{badiaprb01}. The $dc$
measurement is easily corrupted by noise, such as vibrations and
$1/f$ electronic noise. Much higher sensitivity of MFM can be
achieved in the $ac$ mode by driving the cantilever at its resonant
frequency by that one reduces environmental noise and increases the
signal to noise ratio using lock-in or frequency modulation
techniques\cite{zhugrutter}. To develop and to solve the inverse
problem in the case of the $ac$ regime with the inclusion of losses
we follow the method developed by Badia in
Ref.[\onlinecite{badiaprb01}].

First, let us assume the tip $(0,0,m_{z})$ to be harmonically driven
at a constant height above a magnetic slab. The cantilever
oscillation amplitude is considered to be constant as well. This
condition is used for dissipation imaging, an imaging technique
which is sensitive to non-conservative interactions
\cite{zhugrutter,grutter}.

We consider a small vertical oscillation of the tip around the point
located at the distance $a$ from the film. Denote by $a'$ the
distance between the tip and the film. Then we have the relation
$a'=a+A' \cos (w_{d}t)$, where $A'$ is the amplitude of the tip
oscillations, which satisfies the condition $A'\ll a$. The quantity
$w_d$ is used for the driving angular frequency of the MFM tip.

Taking into account the nondispersive ohmic relation
$\vec{J}=\sigma_{n} \vec{E}$ and the well-known time-dependent
Maxwell equations, we get the following relation for the magnetic
field inside the film $\nabla^{2}\vec{B}=\bar{\mu}\sigma_{n}
\partial\vec{B}/\partial t$. The quantity $\sigma_{n}$ is the conductivity and $\bar{\mu}$ is the permeability
of the material.

The relative permeability $\mu_{r}$ to be determined is related to
the real part $\mu'$ of $\bar{\mu}$ as follows:
$\mu'=\mu_{r}\mu_{0}$, where $\bar{\mu}=\{\mu'+i\mu''\}$. Now we are
going to determine $\bar{\mu}$ as a first step in order to get
$\mu_{r}$. To that end, we shall find the force acting on the tip
and solve the inverse problem.

Replacing $a$ by $a'$ in (\ref{eq:NonLOCAL}), we get
\begin{eqnarray}
\label{eq:alternoa} B_{z1} &=& \frac{\mu_{0} m_{z}}{4
\pi}\int_{0}^{\infty}dk k^{2} e^{-k(a-z)} J_{0}(k\rho)\nonumber
\\
&-& \frac{\mu_{0} m_{z}}{4 \pi} A' \cos (w_{d}t)\int_{0}^{\infty}dk
k^{3} e^{-k(a-z)}
J_{0}(k\rho)\nonumber \nonumber \\
&+& \frac{\mu_{0} m_{z}}{8 \pi} A'^{2}
\cos^{2}(w_{d}t) \int_{0}^{\infty}dk k^{4} e^{-k(a-z)} J_{0}(k\rho).\nonumber \\
\end{eqnarray}
The following similar expressions can be obtained for $B_{zi}$
$(i=2,3,4)$:
\begin{eqnarray}
\label{eq:alternob} B_{z2} &=& \frac{\mu_{0} m_{z}}{4
\pi}\int_{0}^{\infty}dk k^{2} e^{-k(a+z)} I_{z2}(k,\mu')
J_{0}(k\rho)\nonumber
\\
&-& \frac{\mu_{0} m_{z}}{4 \pi} A' \cos (w_{d}t)\int_{0}^{\infty}dk
k^{3} e^{-k(a+z)} I_{z2}(k,\mu')\nonumber \\ &\times&
J_{0}(k\rho)+\frac{\mu_{0} m_{z}}{8 \pi} A'^{2} \cos^{2}(w_{d}t)
\int_{0}^{\infty}dk k^{4} e^{-k(a+z)}\nonumber \\ &\times&
I_{z2}(k,\mu') J_{0}(k\rho),
\end{eqnarray}
\begin{eqnarray}
\label{eq:alternoc} B_{z3} &=& \frac{m_{z}}{4
\pi}\int_{0}^{\infty}dk k^{2} [I^{+}_{z3}(k,\mu') e^{k
z}+I^{-}_{z3}(k,\mu') e^{k z}] e^{-k a}\nonumber \\ &\times&
J_{0}(k\rho)- \frac{m_{z}}{4 \pi} A' \cos(w_{d}t)\int_{0}^{\infty}dk
k^{3} [I^{+}_{z3}(k,\mu') e^{k z}\nonumber \\ &+& I^{-}_{z3}(k,\mu')
e^{k z}] e^{-k a} J_{0}(k\rho)+ \frac{m_{z}}{8 \pi} A'^{2}
\cos^{2}(w_{d}t)\nonumber \\ &\times& \int_{0}^{\infty}dk
k^{4} [I^{+}_{z3}(k,\mu') e^{k z}+I^{-}_{z3}(k,\mu') e^{k z}] \nonumber \\
&\times& e^{-k a} J_{0}(k\rho),
\end{eqnarray}
\begin{eqnarray}
\label{eq:alternod} B_{z4} &=& \frac{\mu_{0} m_{z}}{4
\pi}\int_{0}^{\infty}dk k^{2} e^{-k(a-z-b)} I_{z4}(k,\mu')
J_{0}(k\rho)\nonumber
\\
&-& \frac{\mu_{0} m_{z}}{4 \pi} A' \cos(w_{d}t)\int_{0}^{\infty}dk
k^{3}
 e^{-k(a-z-b)} I_{z4}(k,\mu')\nonumber \\ &\times&
J_{0}(k\rho)+ \frac{\mu_{0} m_{z}}{8 \pi} A'^{2} \cos^{2}(w_{d}t)
\int_{0}^{\infty}dk k^{4}  e^{-k(a-z-b)}\nonumber \\ &\times&
I_{z4}(k,\mu') J_{0}(k\rho).
\end{eqnarray}
Notice that the coefficients $I_{zi}(k,\mu')$(see
Eq.(\ref{eq:Coefficien})) are similar to the ones found in the
magnetostatic case\cite{delacruzPhysB04}.

In order to find the coefficients $I^{\pm}_{zi}(k,\mu')$, we
consider the following representation of the incident, reflected,
penetrating and transmitted fields as
$B_{zi}(t)=B^{dc}_{zi}+B^{w_{d}}_{zi}(w_{d},t)+B^{2w_{d}}_{zi}(w_{d},t)$,
where $(i=1,\ldots,4)$.

Let us denote by ${\textsl{w}}$ a formal variable which takes the
values $w_{d}$ and $2w_{d}$. Assuming the notation
$B^{ac}$=Re$[\tilde{B}e^{i{\textsl{w}}t}]$ and taking into account
the time-independent solutions (\ref{eq:NonLOCAL}-\ref{eq:Lola2})
for $B^{dc}_{zi}$ (i=2,3,4), we get the following governing
differential equations
\begin{eqnarray}
\label{eq:NaBLa}
\nabla^{2}\tilde{B}^{\textsl{w}}_{z2}(t) &=& 0, \nonumber \\
\nabla^{2}\tilde{B}^{\textsl{w}}_{z3}(t) &=& i2\delta^{-2}(\textsl{w})\,\tilde{B}^{\textsl{w}}_{z3}(t), \\
\nabla^{2}\tilde{B}^{\textsl{w}}_{z4}(t) &=& 0 \nonumber,
\end{eqnarray}
where the skin depth $\delta(\textsl{w})$ is defined by
$\delta(\textsl{w})\equiv(2/\textsl{w} \,\bar{\mu}\,
\sigma_{n})^{1/2}$.

We note that Eq.(\ref{eq:NaBLa}) for $\tilde{B}^{\textsl{w}}_{3}$ is
analogous the equation $\nabla^{2}\vec{B}=(1/\lambda^{2})\vec{B}$,
which arises from the London equation for the superconductor
($\lambda$ is the London penetration depth).

In a similar way to Eqs.(\ref{eq:NonLOCAL}-\ref{eq:Lola2}), we find
solutions $B^{w_{d}}_{zi}$ and $B^{2w_{d}}_{zi}$ $(i=2,3,4)$ of
system (\ref{eq:NaBLa}) as well as $B^{dc}_{zi}$ in the form:
\begin{eqnarray}
\label{eq:alterno} B^{dc}_{z2} &=& \frac{\mu_{0} m_{z}}{4
\pi}\int_{0}^{\infty}dk\,k^{2} e^{-k(a+z)} I_{z2}(k,\mu')
J_{0}(k\rho)\nonumber
\\
&+& \frac{\mu_{0} m_{z}}{16 \pi} A'^{2} \int_{0}^{\infty}dk\,k^{4}
e^{-k(a+z)} I_{z2}(k,\mu') J_{0}(k\rho),\nonumber \\
\end{eqnarray}
\begin{eqnarray}
\label{eq:alterno} \tilde{B}^{w_{d}}_{z2} &=& -\frac{\mu_{0}
m_{z}}{4 \pi} A' \int_{0}^{\infty}dk\,k^{3} e^{-k(a+z)}
\widetilde{{\cal I}}^{w_{d}}_{z2}(k,\mu')
J_{0}(k\rho), \nonumber \\
\end{eqnarray}
and
\begin{eqnarray}
\label{eq:Get back} \tilde{B}^{2w_{d}}_{z2} &=& \frac{\mu_{0}
m_{z}}{16 \pi} A'^{2} \int_{0}^{\infty}dk\,k^{4} e^{-k(a+z)}
\widetilde{{\cal
I}}_{z2}^{2w_{d}}(k,\mu') J_{0}(k\rho)\:.\nonumber \\
\end{eqnarray}
The corresponding expressions for the penetrated fields are:
\begin{eqnarray}
\label{eq:alterno} B^{dc}_{z3} &=& \frac{m_{z}}{4
\pi}\int_{0}^{\infty}dk\,k^{2} [I_{z_{3}}^{+}(k,\mu') e^{k
z}+I_{z_{3}}^{-}(k,\mu') e^{-k z}] \nonumber \\&\times& e^{-ka}
J_{0}(k\rho)\nonumber
\\
&+& \frac{m_{z}}{16 \pi} A'^{2} \int_{0}^{\infty}dk\,k^{4}
[I_{z3}^{+}(k,\mu') e^{k z}+I_{z3}^{-}(k,\mu') e^{-k z}]\nonumber
\\&\times& e^{-ka}
J_{0}(k\rho),
\end{eqnarray}
and
\begin{eqnarray}
\label{eq:alterno} \tilde{B}^{w_{d}}_{z3}(k,\mu') &=&
-\frac{m_{z}}{4 \pi} A' \int_{0}^{\infty}dk\,k^{2} [\widetilde{{\cal
I}}_{z3}^{w_{d}+}(k,\mu') e^{k z}\nonumber \\&+& \widetilde{{\cal
I}}_{z3}^{w_{d}-}(k,\mu') e^{-k z}]e^{-ka} J_{0}(k\rho),\\
\tilde{B}^{2w_{d}}_{z3}(k,\mu') &=& \frac{m_{z}}{16 \pi} A'^{2}
\int_{0}^{\infty}dk\,k^{4} [\widetilde{{\cal
I}}_{z3}^{2w_{d}+}(k,\mu') e^{k z}\nonumber \\&+& \widetilde{{\cal
I}}_{z3}^{2w_{d}-}(k,\mu') e^{-k z}] e^{-ka} J_{0}(k\rho),
\end{eqnarray}
and the transmitted fields are given by
\begin{eqnarray}
\label{eq:alterno} B^{dc}_{z4}(k,\mu') &=& \frac{\mu_{0} m_{z}}{4
\pi}\int_{0}^{\infty}dk\,k^{2} e^{-k(a-z-b)} I_{z4}(k,\mu')
J_{0}(k\rho)\nonumber
\\
&+& \frac{\mu_{0} m_{z}}{16 \pi} A'^{2} \int_{0}^{\infty}dk\,k^{4}
e^{-k(a-z-b)} I_{z4}(k,\mu')\nonumber \\ &\times& J_{0}(k\rho),
\end{eqnarray}
and
\begin{eqnarray}
\label{eq:alterno} \tilde{B}^{w_{d}}_{z4}(k,\mu') &=& -\frac{\mu_{0}
m_{z}}{4 \pi} A' \int_{0}^{\infty}dk\,k^{2} e^{-k(a-z-b)}
\widetilde{{\cal I}}^{w_{d}}_{z4}(k,\mu')\nonumber \\ &\times&
J_{0}(k\rho), \\
\tilde{B}^{2w_{d}}_{z4}(k,\mu') &=& \frac{\mu_{0}
m_{z}}{16 \pi} A'^{2} \int_{0}^{\infty}dk\,k^{4} e^{-k(a-z-b)}
\widetilde{{\cal I}}_{z4}^{2w_{d}}(k,\mu')\nonumber \\ &\times&
J_{0}(k\rho).
\end{eqnarray}

To obtain the coefficients $\widetilde{{\cal
I}}^{w_{d}}_{z2}(k,\mu')$ and $\widetilde{{\cal
I}}_{z2}^{2w_{d}}(k,\mu')$, we must impose the continuity boundary
conditions  on each field components at the interfaces $z=0$ and
$z=-b$.

Now we shall derive the time-dependent force acting on the
oscillating tip. In fact, in the dipole limit, the instantaneous
value of this force may be calculated from the expression ${\cal
F}_{z}(t)=-\partial_{a'}\{{-(1/2)\vec{m}\cdot
\vec{B}_{2}[a'(t)]}\}$.

The magnetic force ${\cal F}_{z}(t)$ can be decomposed into the sum
${\cal F}_{z}(t)=F^{dc}_{z}+F_{z}^{ac}(t)$. The linear approximation
with respect to $A'$ for the components of ${\cal F}_{z}(t)$ can be
written as follows
\begin{figure}[!tbp]
\centerline{
\includegraphics[width=3.2in]{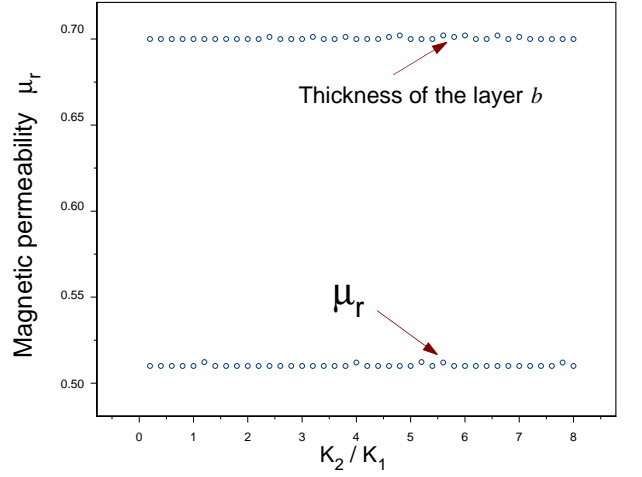}
} \caption{The graph shows the real part of the magnetic
permeability $\mu_{r}$ and the thickness of the film $b$ by HIM
applied to ac MFM.} \label{figure5}
\end{figure}
\begin{eqnarray}
\label{eq:alterno} F^{dc}_{z} &=& \frac{\mu_{0} m^{2}_{z}}{8 \pi}
\int_{0}^{\infty}dkk^{3} e^{-2ka} I_{z2}(k,\mu')J_{0}(k\rho) \nonumber \\
&+& \frac{\mu_{0} m^{2}_{z}}{8 \pi} \int_{0}^{\infty}dkk^{3}
e^{-2ka} |\widetilde{{\cal I}}^{w_{d}}_{z2}|(k,\mu')\nonumber \\
&\times& J_{0}(k\rho)\cos{\varphi}^{w_{d}},
\end{eqnarray}
and
\begin{eqnarray}
\label{eq:altern} F^{ac}_{z} &\simeq& -\frac{\mu_{0} m^{2}_{z}}{8
\pi}
\int_{0}^{\infty}dk\,k^{4} e^{-2ka} {I}_{z2}(k,\mu')J_{0}(k\rho)\nonumber \\ &\times& A'\cos({w_{d}t})\nonumber \\
&-& \frac{\mu_{0} m^{2}_{z}}{4 \pi} \int_{0}^{\infty}dkk^{4}
e^{-2ka} |\widetilde{{\cal I}}^{w_{d}}_{z2}|(k,\mu')
\,J_{0}(k\rho)\nonumber \\ &\times& A'\cos({w_{d}t+\varphi^{w_{d}}})
\nonumber
\\ &-& \frac{\mu_{0} m^{2}_{z}}{8 \pi} \int_{0}^{\infty}dkk^{4} e^{-2ka}
|\widetilde{{\cal I}}^{2w_{d}}_{z2}|(k,\mu')J_{0}(k\rho)\nonumber \\
&\times& A'\cos({2w_{d}t+\varphi^{2w_{d}}}),
\end{eqnarray}
were $\widetilde{{\cal I}}^{w_{d}}_{z2}=|\widetilde{{\cal
I}}^{w_{d}}_{z2}|e^{i\varphi^{w_{d}}}$ and $\widetilde{{\cal
I}}^{2w_{d}}_{z2}=|\widetilde{{\cal
I}}^{2w_{d}}_{z2}|e^{i\varphi^{2w_{d}}}$ and $\varphi^{w}$ is the
phase lag between the tip-sample force and the displacement of the
tip.

In the lowest-order approximation to the $ac$ problem, we can assume
that $\varphi^{w}$ is a small parameter\cite{badiaprb01}.

Taking into account that the force is recovered in the form
$F^{ac}_{z}=\mathrm{Re}[\tilde{F}_{z}e^{iw_{d}t}]$, we can rewrite
the Eq.(\ref{eq:altern}) as follows
\begin{equation}
\label{eq:alterno} \tilde{F}_{z}=-\frac{\mu_{0} A'  m^{2}_{z}}{2
\pi} \int_{0}^{\infty}dkk^{4} e^{-2ka} {\cal I}_{z2}(k,\mu',w_{d})
J_{0}(k\rho),
\end{equation}
where we define ${\cal I}_{z2}(k,\mu',w_{d})\equiv
(1/4)I_{z2}(k,\mu')+(1/2)\widetilde{{\cal
I}}^{w_{d}}_{z2}(k,\mu',w_{d})+(1/4)\widetilde{{\cal
I}}^{2w_{d}}_{z2}(k,\mu',2w_{d})$.

Now, we shall recover $\mu_{r}$ and $b$ from MFM measurements in
$ac$ modes. First, recall that the complex amplitude of the $ac$
force
$\tilde{F}_{z}=\mathrm{Re}[\tilde{F}_{z}]+i\mathrm{Im}[\tilde{F}_{z}]$
may be experimentally found.

Here, we show that a straightforward extension of the vector
inversion method to complex variables can be done.
Eq.(\ref{eq:alterno}) can be written using the operator $\ScH_{0}$
as follows
\begin{equation}
\label{eq:secondpart} \tilde{F}_{z}(a_0,\rho)=-\frac{\mu_{0} A'
m^{2}_{z}}{2 \pi}\ScH_{0}[k^{3} e^{-2ka_{0}}{\cal
I}_{z2}(k,\mu',w_{d})].
\end{equation}
Then, we can perform the mathematical inversion for the pairs
$[\rho_{i},\tilde{F}_{z}(a_{0},\rho_{i})]$. In particular, one can
apply the inverse operator $\ScH_{0}^{-1}$ to both sides of
Eq.(\ref{eq:secondpart}) and obtain the following system for any
$k_{1}\neq k_{2}$
\begin{eqnarray}
\label{eq:Kerry}
\frac{\mathcal{G}(k_{1})}{k_{1}^3}\;\ScH_{0}^{-1}[\tilde{F}_{z}(a_0,\rho)](k_{1})+{\cal I}_{z2}(\mu',b;k_{1}) &=& 0 \nonumber \\
\frac{\mathcal{G}(k_{2})}{k_{2}^3}\;\ScH_{0}^{-1}[\tilde{F}_{z}(a_0,\rho)](k_{2})+{\cal
I}_{z2}(\mu',b;k_{2}) &=& 0,
\end{eqnarray}
where $\mathcal{G}(k)\equiv e^{2ka_{0}}\,{2 \pi}/\mu_{0} A'
m^{2}_{z}$.

To solve the system (\ref{eq:Kerry}) for the quantities
$(\mu_{r},b)$, we use the vector inversion procedure together with
the Muller's method\cite{muller}. The Muller's method can be used to
find zeros of a function and can be applied to complex value
functions\cite{badiaprb01}. This method is a generalization of the
secant method in the sense that it does not require the derivative
of a function.

Fig.4 displays the results of recovering the magnetic permeability
and the thickness of the film. These quantities were considered
constants in the process of simulation using noise corrupted fake
data.

\subsection{Frequency oscillation inverse method}

In this section we describe the method of solving the inverse
problem in terms of the shift frequency.

It is convenient to develop the inverse method using the observed
frequency shift $\triangle f$ of the oscillating cantilever rather
than the magnetic force, acting on the tip. If the tip oscillation
amplitude $A'$ is small compared to the distance between the tip and
the slab $a_{0}$, the relation between the gradient of the force
$\partial \tilde{F}_{z}(z)/\partial z$ and $\triangle f$ is given
by\cite{roseman}
\begin{eqnarray}
\label{eq:germany} \frac{\triangle
f}{f_{0}}=\frac{1}{2k_{s}}\frac{\partial \tilde{F}_{z}(z)}{\partial
z},
\end{eqnarray}
where $f_{0}$ is the unperturbed resonance frequency and $k_{s}$ is
the spring constant of the force sensor.

We can rewrite the expression for the force given by
Eq.(\ref{eq:alterno}) in the form
\begin{eqnarray}
\label{eq:alternomodi} \tilde{F}_{z}(z) &=& -\frac{\mu_{0} A'
m^{2}_{z}}{2 \pi} \int_{0}^{\infty}dkk^{4} e^{-k(z+a)} {\cal
I}_{z2}(k,\mu',w_{d})\nonumber \\ &\times& J_{0}(k\rho).
\end{eqnarray}

In the case of the horizontal displacement of the $z-$direction
oscillating tip, Eq.(\ref{eq:germany}) can be written as
\begin{eqnarray}
\label{eq:gurmany} \triangle f &=& \frac{\mu_{0} A'
m^{2}_{z}f_{0}}{4 \pi k_{s}} \int_{0}^{\infty}dkk^{5} e^{-k(z+a)}
{\cal I}_{z2}(k,\mu',w_{d})J_{0}(k\rho).\nonumber \\
\end{eqnarray}
\begin{figure}[!tbp]
\centerline{
\includegraphics[width=3.2in]{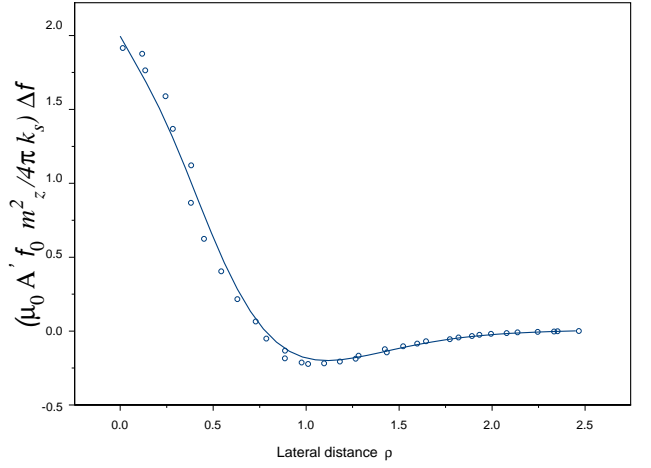}
} \caption{Plot of the simulated frequency shift $\Delta f$ versus
$\rho$. A noise corrupted data have been added.} \label{figure1}
\end{figure}

Finally, applying the operators $\ScH_{0}$ and $\ScH_{0}^{-1}$ to
the Eq.(\ref{eq:gurmany}) and considering $z=a$ and $k$ as
parameters, we get the following system of equations:
\begin{eqnarray}
\label{eq:guurmany} \frac{4\pi k_{s}f_{0}k_{i}^{4}}{\mu_{0} A'
m^{2}_{z}}\,\ScH_{0}^{-1}[\triangle f(a)](k_{i})-{\cal
I}_{z2}(k_{i},\mu',w_{d})=0,
\end{eqnarray}
where $i=1,2$.

We have arrived to a nonlinear system analogous to (\ref{eq:Kerry}),
but this time it is written in terms of the frequency shift $\Delta
f$. The system (\ref{eq:guurmany}) can be solved in the same way as
system (\ref{eq:Kerry}).
\begin{figure}[!tbp]
\centerline{
\includegraphics[width=3.2in]{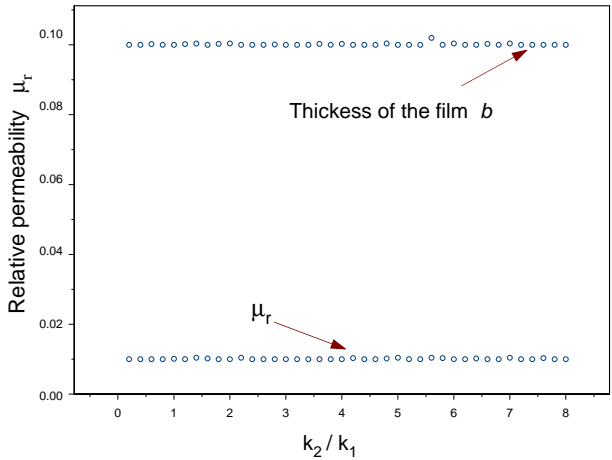}
} \caption{Plot of an example simulated force $F_{z}(a_{0},\rho)$
versus $\rho$ used in this work. A noise corrupted data have been
added.} \label{figure1}
\end{figure}

Fig.5 shows the theoretical shift frequency $\Delta f$ (continuous
line) as well as a simulated measurement (symbols). The values of
$a=0.8,b=0.1,\mu_{r}=0.01$ and $w_{d}=10^{3}rad/s$ were used for our
faked data.
Fig.6 displays results of recovering the magnetic permeability and
the thickness of the film. These quantities were considered constant
in the process of simulation using noise corrupted data.

\section{Conclusion}
Throughout this work, we have developed a theoretical background for
the recovery of the  magnetic permeability in magnetic films by MFM
experiments. In addition, it has been shown that $b$ may be
recovered from the measurements of the vertical force in the case of
a horizontal scanning of the MFM tip.

Finally, we have suggested a solution of the inverse problem using
the shift frequency of the system.

\section{Acknowledgement}
A. de la Cruz thanks to A. Badia and V. Shramshenko. This work was
supported by the Department of Mathematics and Statistics and by the
Physics Department of Concordia University.

\end{document}